\pdfoutput=1
\documentclass[10pt, hidelinks, conference, compsocconf]{IEEEtran}
\ifCLASSINFOpdf
\else
\fi
\usepackage{amsmath}
\usepackage{array}

\usepackage[tight,footnotesize]{subfigure}

\usepackage{caption}
\usepackage{url}

\usepackage{tikz}
\usepackage[nolist]{acronym}
\usepackage{float}
\usepackage{tabularx}
\usepackage{color}															 
\usepackage{colortbl}
\usepackage{booktabs}
\usepackage{multirow}
\usepackage{listings}
\usepackage{amssymb}
\usepackage{mathtools}
\usepackage{scalerel}
\usepackage{balance}

\def\labelitemi{--}

\usetikzlibrary{svg.path}

\definecolor{orcidlogocol}{HTML}{A6CE39}
\tikzset{
  orcidlogo/.pic={
    \fill[orcidlogocol] svg{M256,128c0,70.7-57.3,128-128,128C57.3,256,0,198.7,0,128C0,57.3,57.3,0,128,0C198.7,0,256,57.3,256,128z};
    \fill[white] svg{M86.3,186.2H70.9V79.1h15.4v48.4V186.2z}
                 svg{M108.9,79.1h41.6c39.6,0,57,28.3,57,53.6c0,27.5-21.5,53.6-56.8,53.6h-41.8V79.1z M124.3,172.4h24.5c34.9,0,42.9-26.5,42.9-39.7c0-21.5-13.7-39.7-43.7-39.7h-23.7V172.4z}
                 svg{M88.7,56.8c0,5.5-4.5,10.1-10.1,10.1c-5.6,0-10.1-4.6-10.1-10.1c0-5.6,4.5-10.1,10.1-10.1C84.2,46.7,88.7,51.3,88.7,56.8z};
  }
}

\begin{document}
%

\title{CAN Radar: Sensing Physical Devices in CAN Networks based on Time Domain Reflectometry}

\begin{acronym}
 \acro{KWP}{Keyword Protocol 2000}
 \acro{UDS}{Unified Diagnostic Services}
 \acro{OBD}{On-board diagnostics}
 \acro{CAN}{Controller Area Network}
 \acro{LIN}{Local Interconnect Network}
\acro{MOST}{Media Oriented Systems Transport}
 \acro{ECU}{Electronic Control Unit}
 \acro{WMA}{Windows Media Audio}
 \acro{TPMS}{Tire Pressure Monitoring System}
 \acro{RPM}{Revolutions Per Minute}
 \acro{VANET}{Vehicular Ad Hoc Network}
 \acro{DoS}{Denial of Service}
 \acro{BEV}{Battery Electric Vehicle}
 \acro{IT}{Information Technology}
 \acro{TP}{Transport Protocol}
 \acro{Nmap}{Network Mapper}
 \acro{IP}{Internet Protocol}
 \acro{OSI}{Open Systems Interconnection}
 \acro{ID}{Identifier}
 \acro{TCP}{Transmission Control Protocol}
 \acro{UDP}{User Datagram Protocol}
 \acro{HTTP}{Hypertext Transfer Protocol}
 \acro{SMTP}{Simple Mail Transfer Protocol}
 \acro{GUI}{Graphical user interface}
 \acro{SID}{Service Identifier}
 \acro{LEV}{Level Identifier}
 \acro{PID}{Parameter Identifier}
 \acro{OEM}{Original Equipment Manufacturer}
 \acro{HV}{High Voltage}
 \acro{BEV}{Battery Electric Vehicle}
 \acro{PKI}{Public key infrastructure}
 \acro{CA}{certification authority}
 \acro{HMI}{Human Machine Interface}
 \acro{PCI}{Protocol Control Information}
 \acro{CAESAR}{Competition for Authenticated Encryption: Security, Applicability, and Robustness}
 \acro{CERT}{Computer Emergency Response Team}
 \acro{API}{Application Programming Interface}
 \acro{IDE}{Identifier Extension}
 \acro{OS}{Operating System}
 \acro{USB}{Universal Serial Bus}
\acro{SOF}{Start of Frame}
\acro{RTR}{Remote Transmission Request}
\acro{IDE}{Identifier Extension}
\acro{DLC}{Date Length Code}
\acro{CRC}{Cyclic Redundancy Checksum}
\acro{DEL}{Delimiter}
\acro{ACK}{Acknowledgement}
\acro{EOF}{End Of Frame}
\acro{r}{reserved}
\acro{PCU}{Pyrotechnic Control Unit}
\acro{PDT}{Device Deployment Tool}
\acro{HAZOP}{Hazard and Operability Study}
\acro{ESCL}{Electronic Steering Column Lock}
\acro{ASIL}{Automotive Safety Integrity Level}
\acro{FTA}{Fault Tree Analysis}
\acro{ISO}{International Organization for Standardization}
\acro{SDL}{Security Development Lifecycle}
\acro{SAE}{Society of Automotive Engineers}
\acro{ATA}{Attack Tree Analysis}
\acro{EVITA}{E-Safety Vehicle Intrusion Protected Applications}
\acro{THROP}{Threat and Operability Analysis}
\acro{SGM}{Security Guide-word Method}
\acro{ICT}{Information and Communication Technology}
\acro{CIA}{Confidentiality, Integrity, and Availability}
\acro{CHASSIS}{Combined Harm Assessment of Safety and Security for Information Systems}
\acro{GPS}{Global Positioning System}
\acro{OEM}{Original Equipment Manufacturer}
\acro{ACL}{Additional Communication Line}
\acro{BCM}{Body Control Module}
\acro{BS}{Block Size}
\acro{CAN}{Controller Area Network}
\acro{EOL}{End-of-life}
\acro{IEEM}{Institute of Energy Efficient Mobility}
\acro{GW}{Gateway}
\acro{SDLC}{Software Development Life Cycle}
\acro{OSSTMM}{Open Source Security Testing Methodology Manual}
\acro{PTES}{Penetration Testing Execution Standard}
\acro{ISSAF}{Information Systems Security Assessment Framework}
\acro{OWASP}{Open Web Application Security Project}
\acro{NIST}{National Institute of Standards and Technology}
\acro{PCB}{Printed Circuit Board}
\acro{TARA}{Threat Analysis and Risk Assessment}
\acro{SEI}{Software Engineering Institute}
\acro{SA}{Security Access}
\acro{NRC}{Negative Response Code}
\acro{AES}{Advanced Encryption Security}
\acro{CSCI-ISCW}{Cyber Defense, and Cyber Security}
\acro{CVE}{Common Vulnerabilities and Exposures}
\acro{CGW}{Central Gateway}
\acro{TDR}{Time Domain Reflectometry}
\acro{Acknowledge}{ACK}
\acro{AUTOSAR}{AUTomotive Open System ARchitecture}
\acro{SecOC}{Secure Onboard Communication}
\acro{IDS}{Intrusion Detection System}
\acro{MSE}{Mean Squared Error}
\acro{IFS}{Intermission Frame Space}
 \end{acronym}




\author{\IEEEauthorblockN{Marcel Rumez\IEEEauthorrefmark{1},
J\"{u}rgen D\"{u}rrwang\IEEEauthorrefmark{1},
Tim Brecht\IEEEauthorrefmark{1},
Timo Steinshorn\IEEEauthorrefmark{1}, \\
Peter Neugebauer\IEEEauthorrefmark{1}, 
Reiner Kriesten\IEEEauthorrefmark{1}, and
Eric Sax\IEEEauthorrefmark{2}}
\IEEEauthorblockA{\IEEEauthorrefmark{1}Karlsruhe University of Applied Sciences, Institute of Energy Efficient Mobility, \\
International University Campus 3, 76646 Bruchsal, Germany} 

\IEEEauthorblockA{\IEEEauthorrefmark{2}Karlsruhe Institute of Technology, Institute for Information Processing Technologies,\\
 Engesserstra{\ss}e 5, 76131 Karlsruhe, Germany}} 

\maketitle

\begin{abstract}
The presence of security vulnerabilities in automotive networks has already been shown by various publications in recent years. Due to the specification of the \ac{CAN} as a broadcast medium without security mechanisms, attackers are able to read transmitted messages without being noticed and to inject malicious messages. In order to detect potential attackers within a network or software system as early as possible, \acp{IDS} are prevalent. Many approaches for vehicles are based on techniques which are able to detect deviations from specified \ac{CAN} network behaviour regarding protocol or payload properties. However, it is challenging to detect attackers who secretly connect to \ac{CAN} networks and do not actively participate in bus traffic. In this paper, we present an approach that is capable of successfully detecting unknown \ac{CAN} devices and determining the distance (cable length) between the attacker device and our sensing unit based on \ac{TDR} technique. We evaluated our approach on a real vehicle network.

\end{abstract}



%
\IEEEpeerreviewmaketitle

\section{Introduction}
\label{sec:introduction}

The automotive industry has been undergoing a major transformation process in recent years~\cite{burkacky.2018.mckinsey}. One topic in this transformation process is the information security of vehicle networks, which have become a major focus of \acp{OEM} and suppliers due several cyber-attacks~\cite{Sommer.2019}, in order to be able to primarily ensure the reliability of interconnected systems. This has already resulted in various security specifications for \ac{AUTOSAR}~\cite{AUTOSAR.2016} and the SAE/ISO 21434~\cite{ISO/SAE.CD.21434} standard which is currently in draft. At the same time, more and more intelligent mechanisms are under development to recognize an attacker at an early stage of the attack. One such intelligent technique for the detection of attacks are \acp{IDS} which can be divided into two types~\cite{Mitchell.2014}. First, techniques that use unusual behaviour or signatures from data streams to infer the presence of attackers~\cite{weber2018hybrid,Hamada.2018}. Second, signature-based techniques that are focused on specific patterns in the network to detect malicious actions. Unfortunately, signature-based techniques are not capable to detect unknown attacks (zero-day) due the missing attack pattern (i.e. the attack signature). Based on the assumption that attack techniques become more and more sophisticated, there is the possibility that attacks can be designed in a way that they are not detectable by pattern techniques. On the other hand, behaviour-based approaches are able to identify such unknown attacks due to consideration of unusual behaviour in the information flow of transmission channels. There are already behaviour-based approaches for the \ac{CAN} bus which use deviations from physical characteristics (voltage, current or frequency) of transmitted messages as identification criteria~\cite{Murvay.2014,choi2018.IDS,kneib2018.scission,cho2017viden}. These approaches require a synchronization property like the \ac{ACK} bit or static sections of the \ac{CAN} frame. Moreover, it is only possible to recognize unknown devices if at least one control unit sends messages which excludes the detection of attackers while the \ac{CAN} bus is in idle mode. 

\textbf{Problem:} Existing approaches can only recognize unknown devices if at least one ECU sends out a message and they are bound to fixed parts of the CAN frame.

\textbf{Approach:} We present an approach which is able to physically determine whether an unwanted participant has connected to a \ac{CAN} network. For the detection of unknown participants, the proposed approach is not limited to the protocol structure which distinguishes it from other techniques. Furthermore, the position of the attacker in the network can be calculated.

\textbf{Contribution:} We show a structured procedure for the selection and combination of techniques, which makes it possible to recognize an unwanted participant on the \ac{CAN} bus. In contrast to other approaches, the proposed sensing technique is able to detect an unwanted participant even if no ECU sends a message (bus is in idle state). This also allows to scan for new participants while the vehicle is sleeping (e.g. vehicle is in parking mode) to identify activities such as adding new participants or removing an existing \ac{ECU} over a certain period of time (e.g. for tuning purposes).  

Furthermore, we demonstrate that the approach is also capable of determining the distance (cable length) between the unwanted participant and our detection unit. In addition to a simulative evaluation, which serves for analysis and evaluation of potential detection techniques, we show an application of the approach on a real vehicle network with several \acp{ECU}. The presented approach can either be used in isolation or integrated into existing \acp{IDS} to improve their detection capabilities.

\section{Related Work}
\label{sec:related_work}

\begin{figure*}[htbp]
	\begin{center}							
		\includegraphics[width=0.7\linewidth]{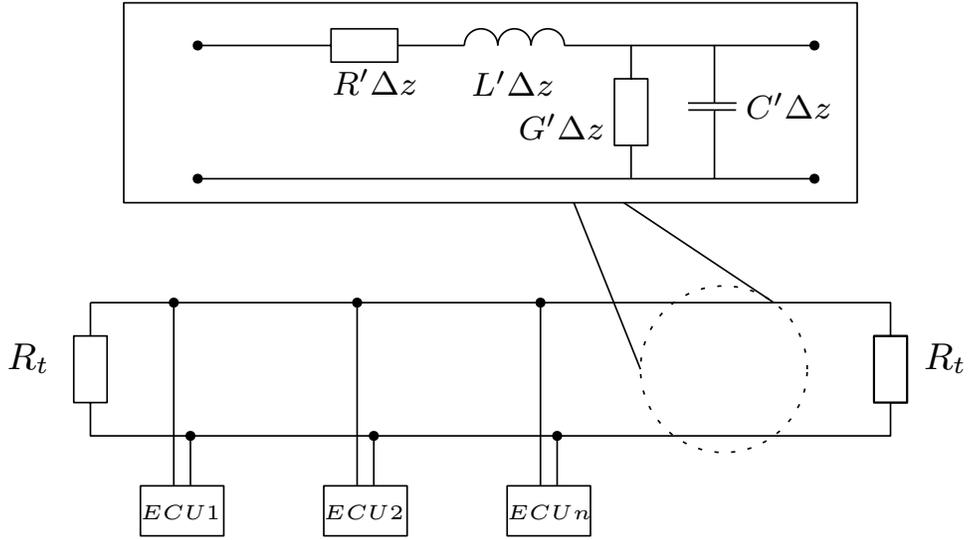}
	\end{center}
	\caption{Typical CAN network with ECUs, termination resistors ($R_t$), corresponding transmission line model with elementary components specified per unit length ($R^{\prime}$ = Distributed resistance, $G^{\prime}$ = Conductance , $L^{\prime}$ = Distributed inductance, $C^{\prime}$ = Capacitance)}
	\label{fig:schaltbild_can2.eps}
\end{figure*}

The authors Choi et al.~\cite{choi2018.IDS} present a method to identify \acp{ECU} by using signal characteristics of transmitted \ac{CAN} messages. They propose different statistical features based on time and frequency for the measured signals. These features are created for each \ac{CAN} message as a kind of fingerprint. Subsequently, an artificial neural network is trained with supervised learning in order to detect deviating characteristics in transmitted \ac{CAN} messages. For the evaluation they used different developer boards as well as corresponding laboratory equipment such as an oscilloscope. In addition, the researchers state a misclassification rate of 0.36\,\%. The limitation of their approach is on the non-detection of pure eavesdropping attacks. A downside of this approach is that the fingerprinting depends on the \ac{CAN} protocol. This leads to the situation that a change of the protocol structure, e.g, as it is necessary with CAN-FD, leads to an adaptation of the evaluation algorithm and thus to a firmware update. Furthermore, the detection algorithm requires that at least one participant sends out a \ac{CAN} message. If this is not the case, an unknown participant cannot be detected. 

Kneib and Huth developed \textit{Scission}~\cite{kneib2018.scission} which also uses physical characteristics (analog values) that belong to the \ac{CAN} frame to recognize a participant. Scission improves the approach from Choi et al.~\cite{choi2018.IDS} due to minimizing resource requirements. The researcher have reduced the sampling rate by a factor of 125 as well as replaced the neural network engine with a logistic regression. The latter allows to model the distribution of dependent discrete variables~\cite{Eyduran.1996}. Thus, the individual bits of the \ac{CAN} frames are analyzed and distributed to three groups. By looking at the individual groups it is then possible to make relevant characteristics more visible. The researchers were able to reduce the false positive, with an average probability for transmitter detection of $99.85\%$. The value was achieved in an evaluation with two production vehicles and a prototype. The system is further able to identify unmonitored and additional devices. As \textit{Scission} is based on the work of Choi et al. and is conceptually comparable (i.e. using the \ac{CAN}-frames), the same drawbacks apply for \textit{Scission}.

A further voltage level based approach for intrusion detection was presented in~\cite{cho2017viden} by the authors Cho and Shin with \textit{Viden}. The detection method uses mainly the ACK-bit voltage threshold of each \ac{CAN} message for creating specific voltage profiles and maps these characteristics on the legitimate \acp{ECU}. During the runtime, the \ac{IDS} update the voltage profiles permanently. If an attack is identified by their system, a voltage profile of the malicious message is created. The profile then allows an assignment of which \ac{ECU} sent the malicious message. The limitations of the method are the following. In case, the attacker \ac{ECU} does not send any messages, no voltage profile can be created for this device. Furthermore, if the attacker uses a valid message with modified payload of an compromised \ac{ECU}, a detection is impossible. Attacks regarding eavesdropping are also out of scope.

\section{Background}
\label{sec:background}
The following sections show the relevant fundamentals regarding the signal transmission and physical properties of the CAN.

\subsection{Transmission Line}
For the transmission of \ac{CAN} messages in vehicles, unshielded twisted pair cables are usually used in vehicles. In transmission-line theory, this type of line is classified in the category of homogeneous twin wires~\cite{Strau.2012}. In Figure~\ref{fig:schaltbild_can2.eps} a typical \ac{CAN} network as well as the transmission line model is shown. It represents an infinite series of fundamental components. The illustrated model shows an infinitely short part of the transmission line. The model includes four elements depending on the their length $\Delta z$:

\begin{itemize}
	\renewcommand{\labelitemi}{$\bullet$}
	\item Distributed resistance $R^{\prime}$ (series of resistors expressed in $\frac{\Omega}{m}$) 
	\item Conductance $G^{\prime}$ (due to the dielectric material of the transmission line expressed in $\frac{S}{m}$)
	\item Distributed inductance $L^{\prime}$ (series of inducers expressed in $\frac{H}{m}$ )
	\item Capacitance $C^{\prime}$ (due to the two transmission lines expressed in $\frac{F}{m}$)
\end{itemize}

The \textit{telegrapher's equations}~\cite{heaviside2008electromagnetic}, defined by Heaviside in the 1880s, describes the current and voltage on such a transmission line depending on distance and time. The equations considering the limit $\Delta z \to$ 0:

\begin{equation}
\frac{\partial I}{\partial z} = -G'V-C'\frac{\partial V}{\partial t}
\end{equation}

\begin{equation}
\frac{\partial V}{\partial z} = -R'I-L'\frac{\partial I}{\partial t}
\end{equation}

This leads to the characteristic impedance $Z_0$:




\begin{equation}
Z_0 = \sqrt{\frac{R' + jwL'}{G' + jwC'}}
\end{equation}
The characteristic impedance is generally a property of the used line, which depends on geometry and material. The assumption as a lossless line ($R^{\prime} \ll jwL^{\prime}$ and $G^{\prime} \ll jwC^{\prime}$) leads to the characteristic impedance becoming real. As a result, the following equation is derived:

\begin{equation}
Z_0 = \sqrt{\frac{L'}{C'}}
\end{equation}

For transmission lines in CAN networks, the ISO 11898~\cite{ISO.12.2015} define cables with a characteristic impedance of $120\,\Omega$. To avoid signal reflections, each transmission line should be equipped with a terminating resistor ($R_{t}$, s. Figure~\ref{fig:schaltbild_can2.eps}), which is equal to $Z_0$. 
\subsection{Time Domain Reflectometry}
\begin{figure}[htbp]
	\begin{center}							
		\includegraphics[width=0.9\linewidth]{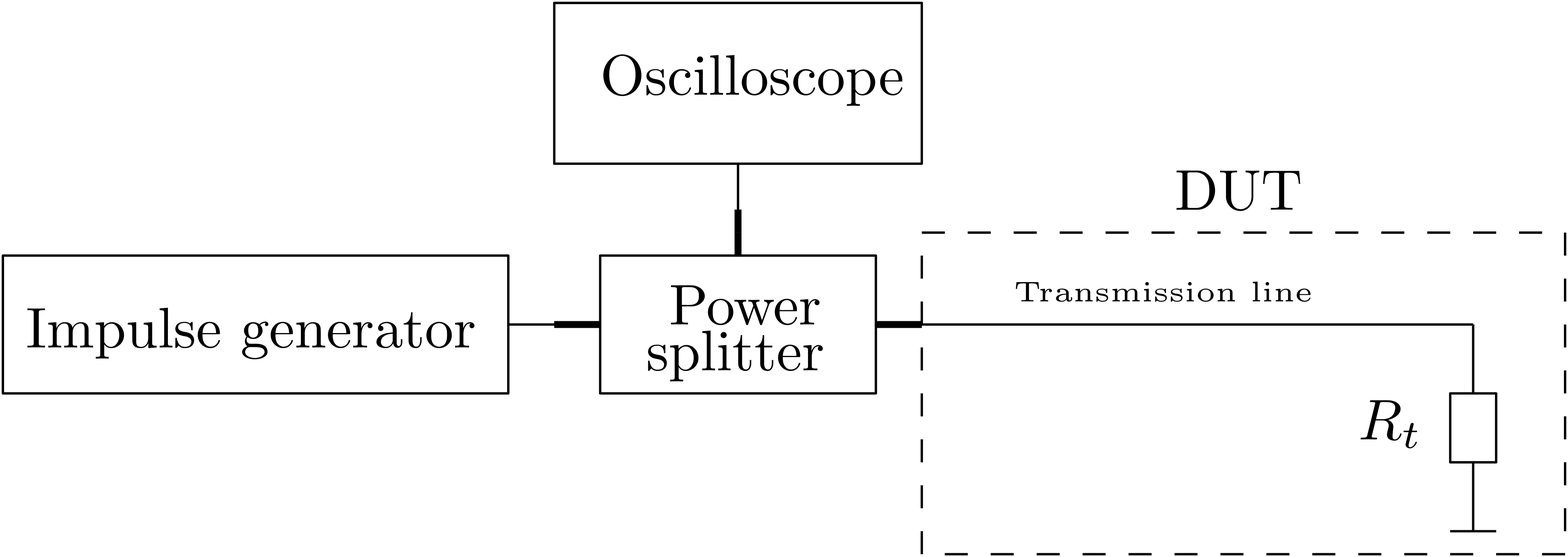}
	\end{center}
	\caption{Setup of TDR measurement technique}
	\label{fig:TDR.eps}
\end{figure}

\ac{TDR}~\cite{Iskander.1992} describes a technique for applying an impulse to a line in order to evaluate the reflected signal, which is often used for detecting errors (e.g. line interruption) in telecommunication lines. A reflection occurs at each impedance jump on the line, which can be caused by stub lines. By determining the runtime of the reflected pulse, the length between measuring point and impedance shift can be calculated. The typical \ac{TDR} structure (s. Figure~\ref{fig:TDR.eps}) includes a pulse generator, oscilloscope and a power splitter.

\section{Approach}
\label{sec:approach}
For the detection of an anonymous device in \ac{CAN} networks, we present our selection process of suitable physical characteristics. As a second step, we present our simulation model to evaluate potential data analysis techniques.

\subsection{Analyzing Suitable Measurement Techniques}
\subsubsection{Ohmic Resistance}
An approach to detect a new participant on the bus is to evaluate the ohmic resistance of the network. Each participant is connected to the bus via a \ac{CAN} transceiver. This transceiver has a specific input resistance. If several of these transceivers are now connected to one bus, the total resistance results from the parallel connection of all input resistors as well as the termination resistors.

A new participant reduces the overall resistance of the network. A typical value for the input resistance of a CAN transceiver is $70\,\text{k}\Omega$. This value is very high compared to those of the termination resistors with $120\,\Omega$. The change in resistance due to the parallel connection of a high-impedance resistor to a low-impedance resistor ($R_{t}$) is only very small. For clarification, an example calculation is given here.
The total resistance for two participants ($R_1$, $R_2$) of the CAN bus with terminating resistance can be calculated using the following equation (without transmission line resistance):

\begin{equation}
R_{tot} =  \left( \frac{1}{R_{t,1}}  + \frac{1}{R_{t,2}} + \frac{1}{R_{1}} + \frac{1}{R_{2}} \right)^{-1}
\end{equation}

With a terminating resistor of $120\,\Omega$ and input resistors of the CAN transceivers ($70\,\text{k}\Omega$)~\cite{TexasInstruments.2018} the result is $R_{ges} = 59.8973\,\Omega$.
If an additional participant would have been added to the network, there would be a minor change of $\Delta R = 0.0512\,\Omega$ which becomes even smaller for systems with many participants.

Furthermore, each CAN transceiver can be considered as a spare voltage source with internal resistance. If several transceivers transmit simultaneously, the voltage sources are connected in parallel with an internal resistance. Depending on how many participants are currently transmitting a recessive level, the voltage on the bus also changes slightly. The \ac{ACK} bit is a state in which all connected participants simultaneously output a recessive level. Using the \ac{ACK} bit for reliable participant recognition is not suitable, since a malicious participant could simply ignore the \ac{ACK} request in order to remain undetected (this drawback applies to \textit{Viden}~\cite{cho2017viden}).

\subsubsection{Impedance}
The impedance measurement uses a network analyzer to measure the input reflection factor in a specific frequency range. According to transmission line theory, the input impedance of the entire bus network should change if an additional device is connected to the bus. This measurement was carried out in a laboratory and showed that impedance changes were hardly measurable when an additional device was attached. 

\subsubsection{Reflection Through Stub Lines}
Another possibility to detect an additional node on the bus is to evaluate the impulse response of the CAN network. Each participant on the \ac{CAN} (whether terminated or not) is represented by a non-ideal terminating resistor. As a result, the impedance changes at each line stub and at the transmitter-receiver side which causes reflections. These reflections can be recorded using \ac{TDR} and evaluated for participant recognition. However, the damping property of the transmission lines could become problematic, since the reflected signals are only very weak and can be lost by passing the line.

\subsubsection{Selection of Suitable Technique}
The first technique has already been investigated and applied by various researchers. The methods are based on changes of voltage and frequency of transmitted signals, which are triggered by variations in the ohmic resistance. In addition, the second measuring technique (impedance) turns out to be unsuitable due to practical lab tests. As a result we decided on the TDR technique, which is already used in other domains for similar purposes (error detection in telecommunication lines). Moreover, TDR offers the advantage of detecting additional devices even if there is no active bus communication.

\subsection{Simulation Model}
The circuit simulation program LTSPICE~\cite{LTSPICE} is used to simulate the CAN network. A simulation model is created which has the structure of a linear bus and both ends are provided with a corresponding terminating resistor. From the actual bus line, short stub lines are routed to the individual participants. The simulation model is used to analyze the effect by adding a participant to the TDR reflection image as well as finding the best analysis method for anomaly detection.

\subsubsection{Transmission Line}
To simulate the CAN line, LTSPICE provides the model of an ideal line (transmission line).
The model has two parameters. One is the characteristic impedance and the other is the delay time required for a wave to pass through the line. The characteristic impedance is based on a twisted two-wire line, which is typically used in vehicles. The RF simulation program QUCS-Studio~\cite{QucsStudio} can be used to calculate the characteristic impedance of a line. The values of Table~\ref{tab:cable_charac} have been used in accordance with the \ac{CAN} standard ISO~11898~\cite{ISO.12.2015}. 

\begin{table}[H]
\centering
	\caption{Cable characteristics according to ISO~11898~\cite{ISO.12.2015}}
	\begin{tabular}{l|l}
		\textbf{Property}  & \textbf{Value}  \\ \toprule
		Cable type					&  Twisted pair cable  \\ 
		Wire diameter  &  0.657~mm  \\ 
		Diameter with isolation  & 1.1~mm   \\ 
	  Length   & 1~m, 100 Twists (twisting)  \\ 
	\end{tabular}
		\label{tab:cable_charac}
\end{table}

\begin{table*}[!ht]
	\centering
	\caption{Comparison of the different analysis techniques: - no change detected, (\checkmark) change partially detected, \checkmark change continuously detected. The top row shows the seven \acp{ECU} (ARS, DME, DSC, eetc. ) of the Powertrain-\ac{CAN} which were used for evaluation of the four analysis techniques.}
	\vspace{0.1cm}
	\begin{tabular}{l|l|l|l|l|l|l|l}
		& \textbf{ARS}  & \textbf{DME} & \textbf{DSC}  & \textbf{EKP}  & \textbf{Light}  & \textbf{Engine}  & \textbf{SZL LWS} \\ \toprule
		Mean Square Error 					&  \checkmark & \checkmark & \checkmark & \checkmark  & \checkmark & - & \checkmark   \\ 
		Correlation Analysis  & \checkmark & \checkmark & \checkmark & (\checkmark) & \checkmark & \checkmark & (\checkmark) \\ 
		RQCC  								& \checkmark  & \checkmark & \checkmark & \checkmark & \checkmark & - &  \checkmark   \\ 
		\rowcolor[gray]{.9} Coherence Analysis & \checkmark & \checkmark & \checkmark & \checkmark & \checkmark & \checkmark & \checkmark   \\ 
	\end{tabular}
	\label{tab:method_comparison}
\end{table*}

The line calculation tool provides a characteristic impedance of approximately $117~\Omega$. In addition, the phase angle (caused by delay) is calculated. This  angle can be converted into a time delay for the LTSPICE model by assuming the typical velocity of prorogation in lines of $ v = 2 \cdot 10^{8}\,\frac{m}{s}$, which results in  $v = \frac{2}{3}\,c$. Since the characteristic impedance of a typical CAN network is different to the impedance of commonly used oscilloscopes and impulse generators, a matching circuit is necessary to suppress the introduction of additional reflections, caused by the measurement equipment.

\subsubsection{CAN Transceiver}
\begin{figure}[htbp]
	\begin{center}
		\includegraphics[width=0.4\linewidth]{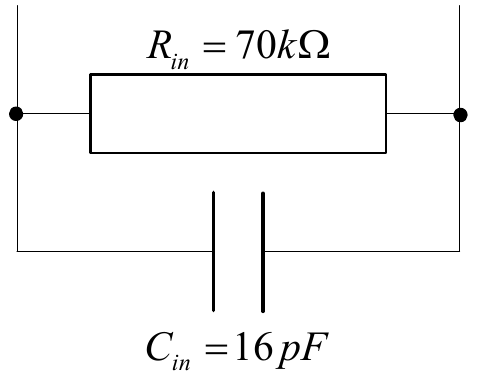}
		\caption{Model of a simulated CAN transceiver, showing the typical internal resistance and capacity.}
		\label{fig:trans_charac}
	\end{center}
\end{figure}

For the simulation of the CAN participants, a simple substitute circuit for emulating the input impedance of a CAN transceiver is created. The equivalent circuit consists of a high-resistance input and a small parasitic input capacity (see Figure~\ref{fig:trans_charac}). 

The typical values from a CAN transceiver data sheet were adopted as values for resistance ($70~\text{k}\Omega$) and capacity ($16~\text{pF}$)~\cite{TexasInstruments.2018}.

\subsection{Data Analysis Techniques} \label{Assessment of Measurement Techniques}
\begin{figure}[htbp]
	\begin{center}
		\includegraphics[width=1.0\linewidth]{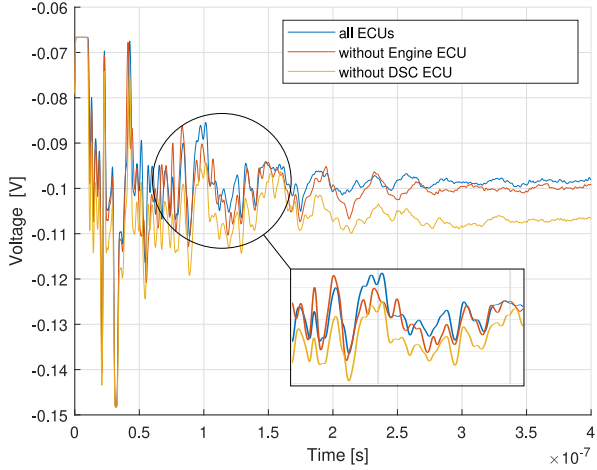}
		\caption{Course of signal reflections by separation of different CAN participants. The blue curve represents $V_{Ref} (t)$ and the other two different $V_{Act,i} (t)$. For details see text. }
		\label{BMW_ref_Motor_DSC_new.eps}
	\end{center}
\end{figure}

In order to detect a variation of the transmission line properties, the reflection behavior of the bus has to be evaluated and continuously compared with a reference signal. The reference signal $V_{Ref} (t)$  represents the normal reflection behavior as voltage over time of the unmodified CAN network. In contrast, the signal $V_{Act} (t)$ refers to the actual measured reflection behavior in the network (see Figure~\ref{BMW_ref_Motor_DSC_new.eps}).  

In Figure~\ref{BMW_ref_Motor_DSC_new.eps}, three signal courses are shown, which represent the reflection behavior due to physical changes in the CAN network. The blue signal can be considered as reference signal since all \acp{ECU} in the network are connected. If several devices get disconnected from the bus, a changing reflection behavior occurs, represented by the red and orange signals, respectively. In order to detect an additional or missing device, it is necessary to compare the previously recorded reference signal with the other signal courses. For this purpose, various data analysis techniques were investigated, which are briefly described below.

\subsubsection{\acf{MSE}}
A simple method to obtain a scale for the difference between two signal courses, is to determine the \ac{MSE}~\cite{Wackerly.2008}. For this purpose, the distance of the measured signal from the reference signal is determined and squared at each point of the respective functions. In this case, the determined distances between the two function curves represent the errors. The method weighs greater differences more strongly by squaring the errors, and eliminates negative signs. The result is the mean value for the error set, where a larger mean square error indicates a larger difference between the two curves. This can be used to determine whether a significant change in the measurement signal is detected and whether an unknown participant is present. The low amount of computation is an advantage of this method. However, an offset error has a negative effect on the calculation. This increases the error at any time, even if the signals show the same changes. Noise also has a significant influence on the result. Since the voltage levels are very low due to the damping of the matching circuit and the power splitter, a voltage peak can make the error large compared to the actual difference. Since the error is squared, a voltage peak caused by noise has a big influence on the result. 

\subsubsection{Cross Correlation Analysis}
Cross-correlation analysis is a generalization of standard linear correlation analysis. It is a method of determining the strength of a relationship between two measured time series~\cite{Rabiner.1978}. If a correlation is found ($r=1$) between two time series, it means that if one time series is changed systematically, the other time series is also changed systematically. Where the value $1$ for the correlation coefficient $r$ means a maximum similarity and the value $0$ means no correlation between the signals~\cite{Cleff.2008}. In this case $r=1$ represents a \ac{CAN} bus without an unknown participant and a value of $r<1$ for a \ac{CAN} bus with an unknown participant.

\subsubsection{Robust Quantitative Comparison Criterion (RQCC)}
Since both the correlation analysis and the quadratic error perform an evaluation in the time domain, methods in the frequency domain were also considered. One such method is the \textit{Robust Quantitative Comparison Criterion}~\cite{Perlin.2014}. It is a robust method to compare two signals by the \textit{Sobolev norm} ($H^S$) in order to determine a difference. The latter presents itself as a distance norm between two functions and contains both phase and amplitude information~\cite{Perlin.2014}. 



\subsubsection{Coherence Analysis} 
\label{Kohaerenzanalyse}
The coherence  assesses the strength of the dependency between the two time signals at a given angular frequency~\cite{White.1990}. Thus, it estimates to which extent the measured signal ($V_{Act}$) of the reference signal ($V_{Ref}$) can be predicted by an optimal linear function of the least squares. The result is a coherence value between $0$ and $1$ where the value $1$ describes the absolute correspondence between the two signals. In our case the value 1 represents no unknown participant on the CAN bus, e.g, $V_{Act} = V_{Ref}$.  

\subsection{Evaluation of Analysis Methods}

For the evaluation of the different analysis methods, the measurements were examined using the simulated vehicle network. In order to obtain a meaningful result, all algorithms use the same signal sections. If the signals are considered over the entire measurement range, only a marginal change in the characteristic quantity is obtained when the network is changed. The latest reflection is limited in time by the length of the CAN bus, which is why a longer observation does not bring any advantage, but only adds disturbances to the evaluation. That is why only a small part of the signal is considered and the averaged signal is compared with the reference value. A part of a certain length is taken from both the averaged signal and the reference signal around the point with the largest difference in magnitude and transferred to the detection technique.

The study of the various methods of analysis showed that the coherence analysis was able to identify all seven \acp{ECU} and thus outperformed the other techniques, which is presented in Table~\ref{tab:method_comparison}. In contrast to the other analysis methods, the output value of the coherence analysis is zero, even if a small physical change (e.g. cable movement) has been made to the network. This can be explained by the detection of peaks. A peak is only detected as such at a certain peak value, since no peaks in the coherence calculation exceed the value, no peak is detected and thus the characteristic value remains zero. 



\subsubsection{Extended Coherence Analysis} 
\label{subsubsec:ex_coherence}
We modified the original coherence analysis with the weighted phase angle to include the phase information. Below we show these modifications of the original approach, which is represented by Equation~\ref{eq:orig_coherence}.

\begin{equation}
\label{eq:orig_coherence}
C(w)=\frac{C_r(w)^2}{f_{x}(w)^2 f_{y}(w)^2}
\end{equation}
With $f_x(w)$ as spectral power density of $x$ and $f_y(w)$ as spectral power density of $y$, the coherence $C(w)$ can reach values of $[0;1]$~\cite{Menges.1982} and where $f_{x}(w)^2$, $f_{y}(w)^2$ are the auto-spectral density of $f_{x}$ and $f_{y}$. Related to our method, $f_{x}$ corresponds to the reference reflections ($V_{ref}(t)$) and $f_{y}$ to the reflections during the active search ($V_{Act}(t)$). Furthermore, $C_r(w)$ represents the cross-spectral density of $f_{x}(w)$ and $f_{y}(w)$~\cite{Randall.1987}. If the signals $x$ and $y$ contain different spectral components, peaks occur in the coherence at the frequencies at which the spectra differs.

Since the phase information is lost through pure coherence, the phase of the cross spectral power density $\varphi_{P_{xy}(e^{jw})}$ is additionally taken into account for this evaluation method.

\begin{equation}
\varphi_{P_{xy}(e^{jw})} = \angle P_{xy}(e^{jw})
\end{equation}
where $P_{xy}(e^{jw})$ is calculated as follows:
\begin{equation}
P_{xy}(e^{jw}) =\sum\limits_{m=-\infty}^\infty R_{xy}^{E}(m) e^{-jwm} 
\end{equation}
with
\begin{equation}
R_{xy}^{E}(m) =\sum\limits_{n=-\infty}^\infty y(n+m) x^*(n)
\end{equation}

Where $P_{xy}(e^{jw})$ is the spectral cross power density~\cite{Vaseghi.2012} and $R_{xy}(m)$ is defined as cross correlation ($x^*$ is the complex conjugate of function $x$)~\cite{Mertins.1996}. The coherence has a value of $C(w)=1$ with identical frequency components. The above-mentioned peaks are now detected and their value weighted with the magnitude of the phase at the frequency position. The sum of the weighted peaks is the final result of our extended coherence analysis.

\begin{equation}
K = \sum (1 - C(w)) |\varphi_{P_{xy}(w)}|
\end{equation}



\section{Evaluation}
\label{sec:eval}

\begin{figure}[htbp]
	\begin{center}
		\includegraphics[width=1.0\linewidth]{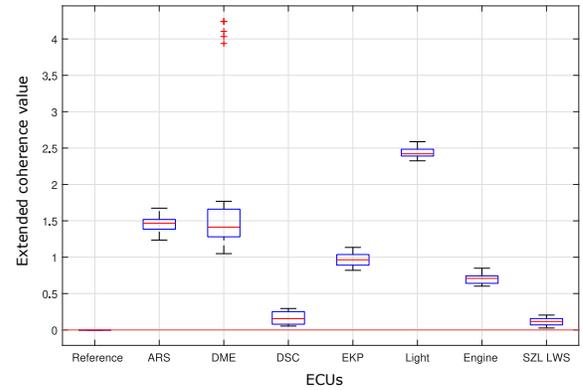}
		\caption{Box plot of the evaluation of measured values with the extended coherence analysis (see above) with physical changes of the CAN. The threshold value is shown as a red line. The red crosses are outliers coming from a source of disturbance.}
		\label{fig:res_cohae}
	\end{center}
\end{figure}

For the  evaluation we carried out the following steps, which would correspond to the sequences in a real application.
\begin{enumerate}
	\item TDR measurement by transmitting a pulse with a width of $3\,ns$ to determine the reference signal $V_{ref}(t)$.
	\item Determine the threshold for detection of an unknown participant.
	\item Start the continuous detection using TDR ($\rightarrow V_{Act}(t)$) and the extended coherence analysis (see Section~\ref{subsubsec:ex_coherence}).
	\item Evaluate the result of the extended coherence analysis and derive the existence of an alien device.
\end{enumerate}
For the detection of physical changes in the observed CAN network, a series of reflections was recorded for calculation of a reference value. At the start of the measurement, all CAN control units are connected to the bus. In order to avoid measurement influences, none of the \acp{ECU} were supplied with power. Subsequently, a control unit was sequentially uncoupled and $1000$ measurements were recorded. This revealed a visible change in the reflection pattern with each pulled off powertrain CAN transceiver (please see the three reflections in Figure~\ref{BMW_ref_Motor_DSC_new.eps}). In this case, even the control unit with the highest distance (dynamic stability control) still showed a visible change of the signal. Furthermore, the results of the measurements showed that when each control unit was disconnected, the measured value was above the defined threshold of $0$ (see Figure~\ref{fig:res_cohae}). Therefore, all changes could be unambiguously detected in this experiment. However, the smallest measured value of the SZL\_LWS-\ac{ECU} shows a value of $0.037$, which is close to the detection limit.

Since \ac{CAN} messages are constantly transmitted on the bus during driving, the recognition has to operate reliably in this state. As a result, the transmitted bits may overlap the measurement. This may influence the reflection pattern, which could lead to incorrect detection. Therefore, the measured signals have to be checked before detecting an attack on an overlay with a message.

\subsection{Attack Scenario with Real Vehicle Network}
\begin{figure}[!htp]  
\begin{center}
		\includegraphics[width=0.9\linewidth]{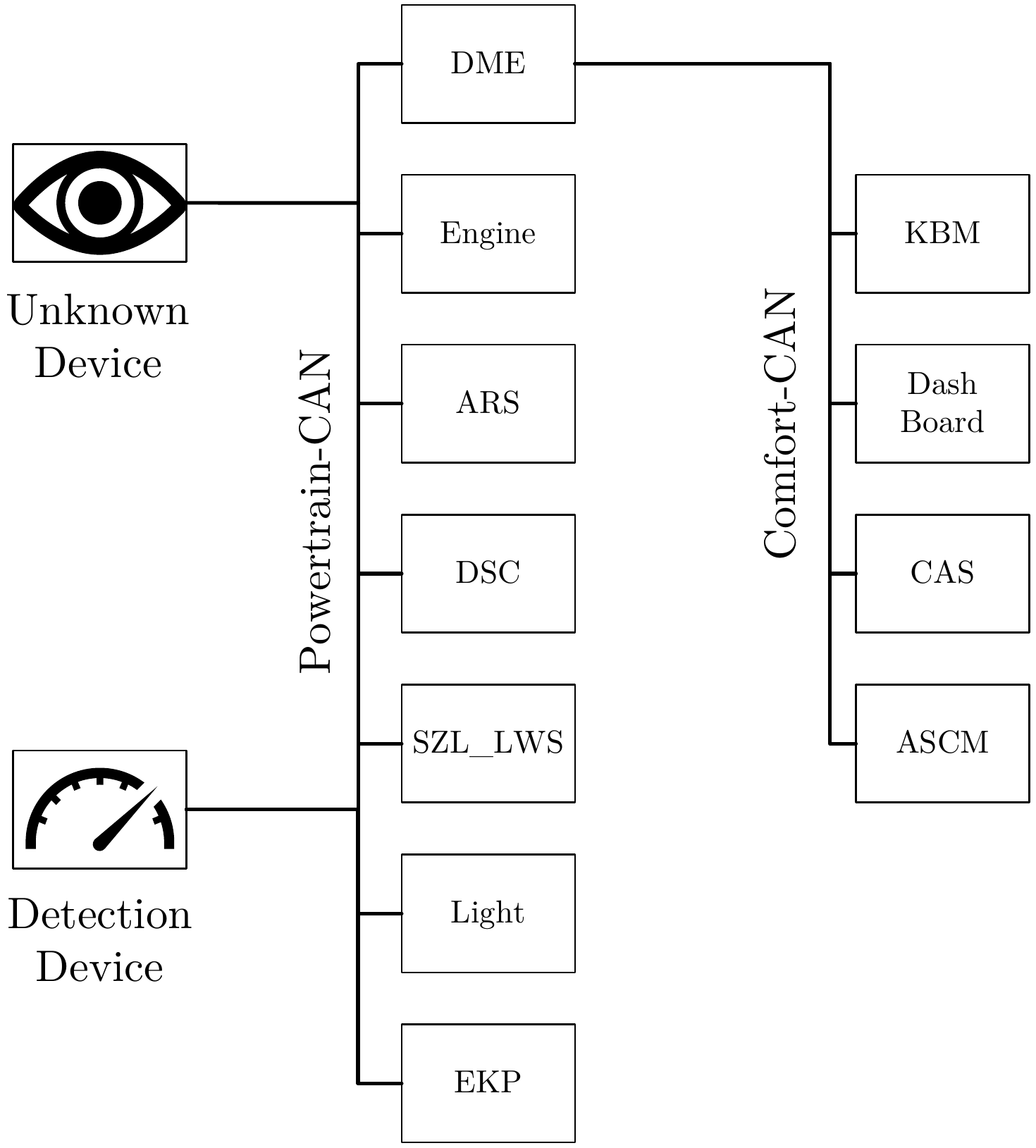}
		\caption{Attack scenario, represented by the Powertrain-CAN with seven connected \acp{ECU} as well as the alien and the detection device.}
		\label{fig:attack_scene}
\end{center}
\end{figure}

To reproduce a possible attack scenario, we used the CAN network shown in Figure~\ref{fig:attack_scene} with predefined cable length shown in Table~\ref{tab:cable_charac}. During the measurement, we added an additional CAN node (Alien Device) to the network. 

\begin{table}[!htp] 
	\centering
	\caption{Distance between \acp{ECU} and detection device}
	\begin{tabular}{l|c}
		\textbf{ECU}  & \textbf{Distance}  \\ \toprule
		EKP					&  8.31~m  \\
		Light					&  5.95~m  \\ 
		SZL\_LWS					&  7.05~m  \\ 
		DSC					&  13.5~m  \\ 
		ARS					&  12.79~m  \\ 
		Engine					&  4.11~m  \\ 
		Unknown Device					& 9.86~m  \\ 
		DME					&  8.65~m  \\ 
	\end{tabular}
	\label{tab:cable_dist}
\end{table}

\begin{figure}[!htp]
	\begin{center}	
			\includegraphics[width=1.0\linewidth]{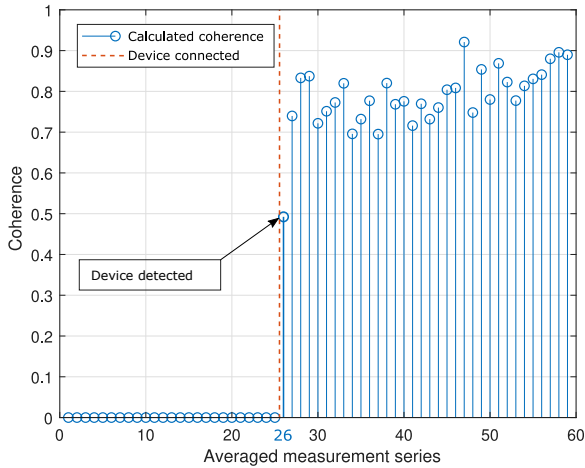}
		\caption{Evaluation of an attack based on coherence calculation}
		\label{Angriffsverlauf_neu.eps}
	\end{center}
\end{figure}

Prior to the attack, 300 measurement series were recorded as reference values ($\overline{V}_{Ref} (t)$). To reduce the noise, 30 measurement series were averaged $\overline{V}_{Act} (t)$ and then compared with the reference signal ($\overline{V}_{Ref} (t)$). Just before the 26th measurement, the unknown device was connected to the CAN bus, which resulted in a significant change in the measurement value at the 26th point (see red dotted line in Figure~\ref{Angriffsverlauf_neu.eps}).


Since both normal and attack values are available for this averaged measurement, the 26th calculated value, as shown in Figure~\ref{Angriffsverlauf_neu.eps}, is not as high as the following. There is however already a clear shift in the coherence value due to the attack, which would be detected with this method. We were also able to determine the exact distance of $9.8~\text{m}$ (see Table~\ref{tab:cable_charac}) between the measuring point and the attacker. The number of averages for defining the reference value is crucial for the evaluation. If too few values are averaged, an attack is sometimes incorrectly detected without changes to the network.


\section{Conclusion and Future Work}
\label{sec:conclusion}
In this work, it was shown that changes in the physical properties of a CAN network can be detected and used for attacker detection. With the combination of a time domain reflectometry and a coherence analysis it is demonstrated that unauthorized CAN participants can be detected reliably. By an evaluation on a real vehicle network, the applicability of the approach has been shown. Since the alien device was successfully detected, an enhanced version of this attack detection mechanism could be used in the automotive industry. In order to avoid possible impairments with regard to the active bus communication, a measurement in vehicle operation could always take place during a transmission stop. For the CAN protocol, an \ac{IFS} is specified, which separates consecutive messages. 

In addition, it would also be conceivable to integrate the method into diagnostic devices in order to locate line interruptions. Audit organizations would also be able to detect additional devices during the general inspection, for example to detect unauthorized tuning devices during the main inspection.

Overall, this method is not limited to CAN networks, but is generally suitable for all networks based on bus technology. This allows the application in large industrial plants using, e.g., ProfiBus or CANopen. 

Further research will focus on the design of a hardware unit suitable for automotive applications, which integrates the oscilloscope, pulse generator and evaluation algorithm. In order to improve the detection technique, a Moving Average Filter could be implemented. This would allow the reference signal to adapt to slow changes in the network properties caused, e.g., by changes in temperature or cable movement.

We would like to point out that our proposed detection technique is not intended to be a replacement for the existent approaches in section~\ref{sec:related_work}, but rather a useful extension of these published techniques.

\section*{Acknowledgements}
This work has been developed in the project AUTO-SIMA (reference number: 13FH006IX6) which is partly funded by the German ministry of education and research (BMBF) within the research programme ICT 2020. The first two authors contributed equally to this work.




\bibliographystyle{IEEEtran}
\bibliography{IEEEabrv,literature}

\end{document}